\documentstyle[aps,prl,preprint,floats,epsfig]{revtex}

\begin{document}

\preprint{\tighten\vbox{\hbox{\hfil CLNS  97/1514}
                        \hbox{\hfil CLEO  97-21}
}}

\title{ Search for Inclusive  $ b \rightarrow s l^+ l^-$ }

\author{CLEO Collaboration}
\date{\today}

\maketitle
\tighten

\begin{abstract}
We have searched for the effective FCNC 
decays  $b\to s l^+ l^- $
using an inclusive method.
We set 
upper limits on the branching ratios 
${\cal B} (b \to s e^+ e^-) < 5.7\times 10^{-5}$,
${\cal B} (b \to s \mu^+ \mu^-) < 5.8\times 10^{-5}$,
and
${\cal B} (b \to s e^{\pm}\mu^{\mp} ) < 2.2\times 10^{-5}$
(at 90 \%\ C.L.).
Combining the di-electron and di-muon decay modes we find:
${\cal B} (b \to s l^+ l^-) < 4.2\times 10^{-5}$ (at 90 \%\ C.L.).
\end{abstract}
\newpage

{
\renewcommand{\thefootnote}{\fnsymbol{footnote}}

\begin{center}
S.~Glenn,$^{1}$ S.~D.~Johnson,$^{1}$ Y.~Kwon,$^{1,}$%
\footnote{Permanent address: Yonsei University, Seoul 120-749, Korea.}
S.~Roberts,$^{1}$ E.~H.~Thorndike,$^{1}$
C.~P.~Jessop,$^{2}$ K.~Lingel,$^{2}$ H.~Marsiske,$^{2}$
M.~L.~Perl,$^{2}$ V.~Savinov,$^{2}$
% ...
\end{center}
 
\small
\begin{center}
$^{1}${University of Rochester, Rochester, New York 14627}\\
$^{2}${Stanford Linear Accelerator Center, Stanford University, Stanford,
California 94309}\\
% ...
\end{center}

\setcounter{footnote}{0}
}
\newpage

Flavor Changing Neutral Currents (FCNC) are forbidden 
to first order in the Standard Model.
Second order loop diagrams, known as penguin
and box diagrams, can generate 
effective FCNC which lead to $b \rightarrow s$ transitions.
These processes are of considerable interest because 
they are sensitive to $V_{ts}$, the Cabibbo-Kobayashi-Maskawa
matrix element
which will be very difficult to measure in direct decays of the 
top quark.
These processes are also sensitive to non-Standard Model physics \cite{nonSM},
since charged Higgs bosons, new gauge bosons or 
supersymmetric particles can contribute via additional loop diagrams.

The electromagnetic penguin decay
$b \rightarrow s \gamma$
was first observed by CLEO in the exclusive mode
$B\to K^*\gamma$ with 
${\cal B}(B \rightarrow K^* \gamma)=(4.2 \pm 0.8 \pm0.6) \times 10^{-5}$ 
\cite{b2kstargammaPRL}. 
The inclusive rate for the decay
$B\to X_s\gamma$ was measured to be
${\cal B}(b\rightarrow s \gamma)=(2.32 \pm 0.57 \pm 0.35)\times 10^{-4}$ 
\cite{b2sgammaPRL}.
Inclusively measured rates are more interesting 
because they can be directly related to underlying 
quark transitions without large theoretical uncertainties in 
formation probabilities for specific hadronic final states.
The measured inclusive $b\to s\gamma$ rate is consistent with Standard
Model calculations.

The  $b \rightarrow s l^+ l^-$  decay rate 
is expected in the Standard Model
to be nearly two orders of magnitude lower than the rate for
$b \rightarrow s \gamma$ decays.
Nevertheless, 
the $b \rightarrow s l^+ l^-$ process has received considerable
attention since it offers a deeper insight into the effective Hamiltonian
describing FCNC processes in $B$ decays \cite{AliMannel}.
While $b \rightarrow s \gamma$ is only sensitive to the
absolute value of the $C_7$  
Wilson coefficient 
in the effective Hamiltonian,
$b \rightarrow s l^+ l^-$ is also sensitive to the sign of $C_7$ and
to the $C_9$ and $C_{10}$ coefficients, 
where the relative contributions vary with $l^+l^-$ mass.
These three coefficients are related to three different processes
contributing to $b \rightarrow s l^+ l^-$: 
electromagnetic and electroweak penguins, and a box diagram.
Processes beyond the Standard Model can alter both the magnitude and the sign
of the Wilson coefficients.
The higher-order QCD corrections for $b\to s l^+l^-$
are  smaller than for the electromagnetic penguin
and have been calculated in next-to-leading order
\cite{refMisiak,refBM}.

Several experiments (UA1 \cite{UA1}, CLEO \cite{CLEOexcl}, and CDF \cite{CDF}) 
have searched for the exclusive decays
$B\to K l^+l^-$ and $B\to K^* l^+l^-$ and set upper limits
at the level of $(1-2)\times 10^{-5}$ at 90\%\ confidence level (C.L.).
Combining electron and muon modes, the previous generation
of the CLEO experiment set an inclusive limit:
${\cal B}(b\rightarrow s l^+l^-)<1.2\times 10^{-3}$ 
(90 \%\ C.L.) \cite{oldCLEO}.
The UA1 experiment \cite{UA1}\ searched for
inclusive $b\rightarrow s\mu^+\mu^-$
at the endpoint of the dilepton mass distribution
($M(\mu^+\mu^-)>3.9$ GeV)
which comprises about a tenth of the total rate.
Extrapolating to the full phase space UA1 claims
a limit of $<5\times10^{-5}$ (90 \%\ C.L.).
However, a simulation of the UA1 acceptance 
shows that UA1 overestimated their efficiency
by at least a factor of three \cite{UA1sim}.

In this Letter, we present results of the search for inclusive
$b\rightarrow s\mu^+\mu^-$, $b\rightarrow s e^+e^-$ 
and $b\to s e^{\pm}\mu^{\mp}$.
The latter decay violates conservation of 
electron and muon lepton numbers
and thus can only originate  from processes beyond the Standard Model.
The data were obtained with the CLEO II detector
at the Cornell Electron Storage Ring.
A sample with an integrated luminosity of 3.1 fb$^{-1}$ was collected on
the $\Upsilon(4S)$ resonance. 
This sample contains $(3.30 \pm 0.06) \times 10^6$
produced $B \bar{B}$ pairs.
For background subtraction 
we also use 1.6 fb$^{-1}$ of data collected just below the  $\Upsilon(4S)$.
CLEO II is a general
purpose solenoidal spectrometer described in detail in Ref.\cite{cleo}.

The data selection method is very similar to the reconstruction method
presented in our previous measurement of the $b\to s\gamma$ rate
\cite{b2sgammaPRL}, with the $\gamma$ candidate replaced by
a lepton pair.
We select events that pass general hadronic event criteria 
based on charged track multiplicity, visible energy and location
of the event vertex.
The highest energy pair of oppositely charged leptons is then selected.
Electron candidates are required to have an energy deposition in the
calorimeter nearly 
equal to the measured momentum, and to have a specific ionization ($dE/dx$) 
in the drift chamber consistent with that expected for an electron.
Muon candidates are identified as charged tracks with matching muon-detector
hits at absorber depths of at least three nuclear interaction lengths.
In the $\mu^+\mu^-$ channel, one muon is required to penetrate at least five
interaction lengths.
We then look for a combination of hadronic particles, denoted $X_s$,
with a kaon candidate and 0-4 pions,
which together with the selected lepton pair satisfy
energy-momentum constraints for the $B$ decay hypothesis, $B\to X_s l^+l^-$.
To quantify consistency with this hypothesis we use:
$$ \chi_B^2=\left(\frac{M_B-5.279}{\sigma_M}\right)^2 + 
            \left(\frac{E_B-E_{beam}}{\sigma_E}\right)^2 $$
where $M_B=\sqrt{E_{beam}^2-P_B^2}$, $E_B$, $P_B$ are the
measured energy and momentum
of the $B$ candidate, and $\sigma_M$, $\sigma_E$ are 
experimental errors on $M_B$ and $E_B$ estimated from the
detector resolution and beam energy spread.
The kaon candidate is a charged track with $dE/dx$ and time of flight (ToF)
consistent with the
kaon hypothesis, 
or a $K^0_S\to\pi^+\pi^-$ candidate identified by a displaced
vertex and invariant mass cut. 
A pion candidate is a charged track with $dE/dx$ and ToF consistent with
the pion hypothesis, or a $\pi^0\to\gamma\gamma$ candidate.
At most one $\pi^0$ is allowed in the $X_s$ combination.
In each event we pick the combination that minimizes overall $\chi^2$, which
includes $\chi_B^2$ together with contributions from $dE/dx$, ToF, and $K^0_S$ 
and $\pi^0$ mass deviations, where relevant.

To suppress continuum background we require the event to have
$H_2/H_0 <0.45$, where $H_i$ are Fox-Wolfram moments \cite{Fox}.
We also require $|\cos\theta_{tt}|<0.8$, where $\theta_{tt}$ is the angle
between the thrust axis of the candidate $B$ and the thrust axis of the
rest of the event. 
To suppress $B\bar{B}$ background we require the mass of the $X_s$ system
to be less than 1.8 GeV.
The dominant $B\bar{B}$ background comes from
two semileptonic decays of $B$ or $D$ mesons, which produce the lepton pair
with two undetected neutrinos.
Since most signal events are expected to have
zero or one neutrino, 
we also require the mass of the undetected system in the event
to be less than 1.5 GeV. 
By excluding the mode with $X_s= K\pi^+\pi^-\pi^0$, we reduce
the expected $B\bar{B}$ background by an additional 21\%\
while reducing the signal efficiency by only 6\%.

\begin{figure}[tbh]   
 \centerline{
  \psfig{file=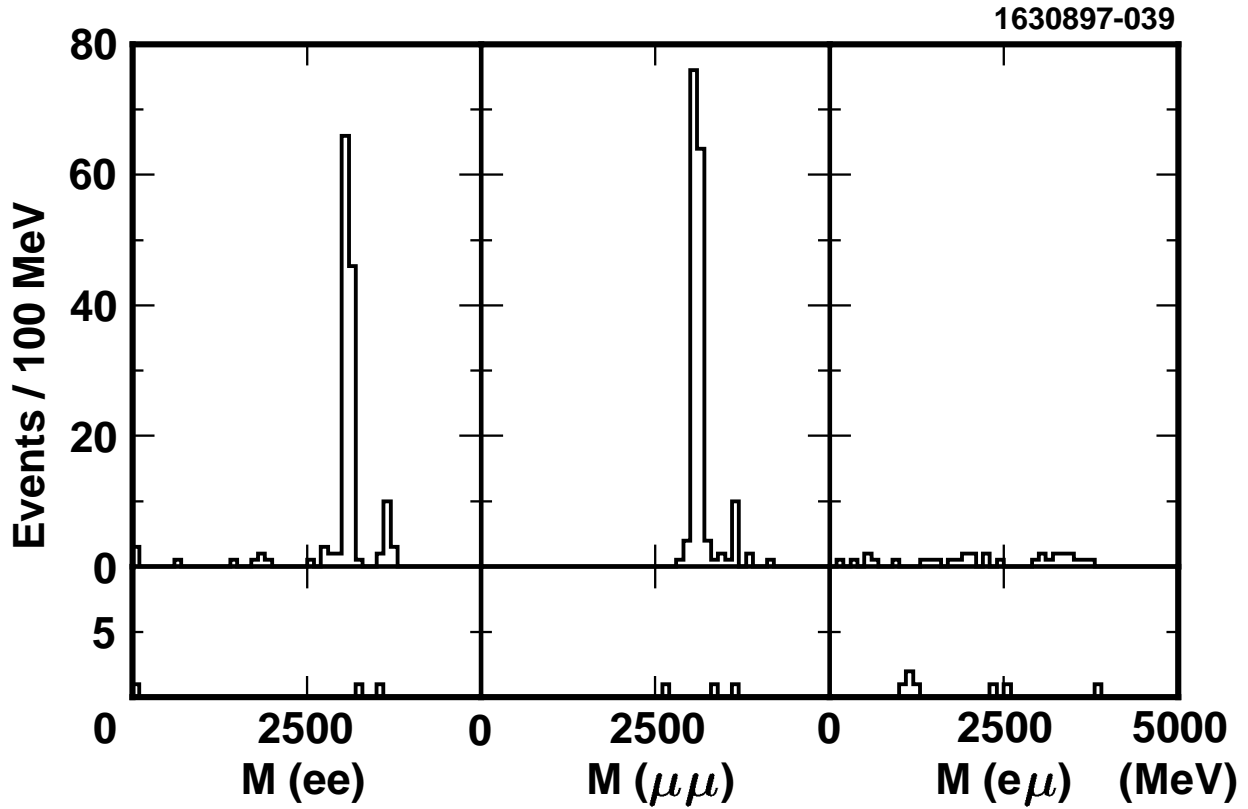,width=\hsize}
 }
\vskip-2.0cm
 \caption{ $M(l^+l^-)$ distributions for the on- (top) and off-resonance
   (bottom) data with the $\chi_B^2<6$ cut. 
  The scaling factor between the off- and on-resonance data is 1.9.    
  \label{fig-mll}
 }
\end{figure}

Fig.~\ref{fig-mll}
shows the dilepton mass, $M(l^+l^-)$,
for the events which pass the cuts previously described and the
$B$ consistency requirement, $\chi_B^2<6$,
in the on- and off-resonance data samples.
Unlike the $b\to s\gamma$ analysis, the continuum background is small.
The peaks at the $\psi$ and $\psi'$ masses that are observed in the 
on-resonance
data are due to the long distance interactions,
$b\to c W^-$, $W^-\to \bar{c}s$, with the $c\bar{c}$ pair 
hadronizing into a $\psi^{(')}$
which then decays to a lepton pair.
Using cuts on $M(X_s)$ to identify $K$ and $K^*$, we measure the
branching ratios for $B\to \psi^{(')}K^{(*)}$
and obtain results consistent with 
a recent CLEO publication \cite{psiCLEO}.
For further analysis, we exclude events with $M(l^+l^-)$ near the
$\psi$ and $\psi'$ masses ($\pm0.1$ GeV for $\mu^+\mu^-$, $-0.3,+0.1$ GeV for
$e^+e^-$, no cut for $e^{\pm}\mu^{\mp}$).
The exclusion region is wider in the $e^+e^-$ channel because of
the radiative tail.
After these cuts and continuum subtraction,
we observe $10\pm5$  $X_s e^+e^-$, 
$12\pm6$ $X_s\mu^+\mu^-$, and $18\pm8$ $X_s e^{\pm}\mu^{\mp}$ 
events in the  data.
From the Monte Carlo simulation of generic $B\bar{B}$ events we expect
$9\pm1$, $16\pm2$ and $39\pm3$ (statistical errors only)
background events, respectively.
Therefore, no evidence for signal is found in the data and we proceed
to set limits on these decay rates.

\begin{figure}[tbhp]   
 \centerline{
  \psfig{file=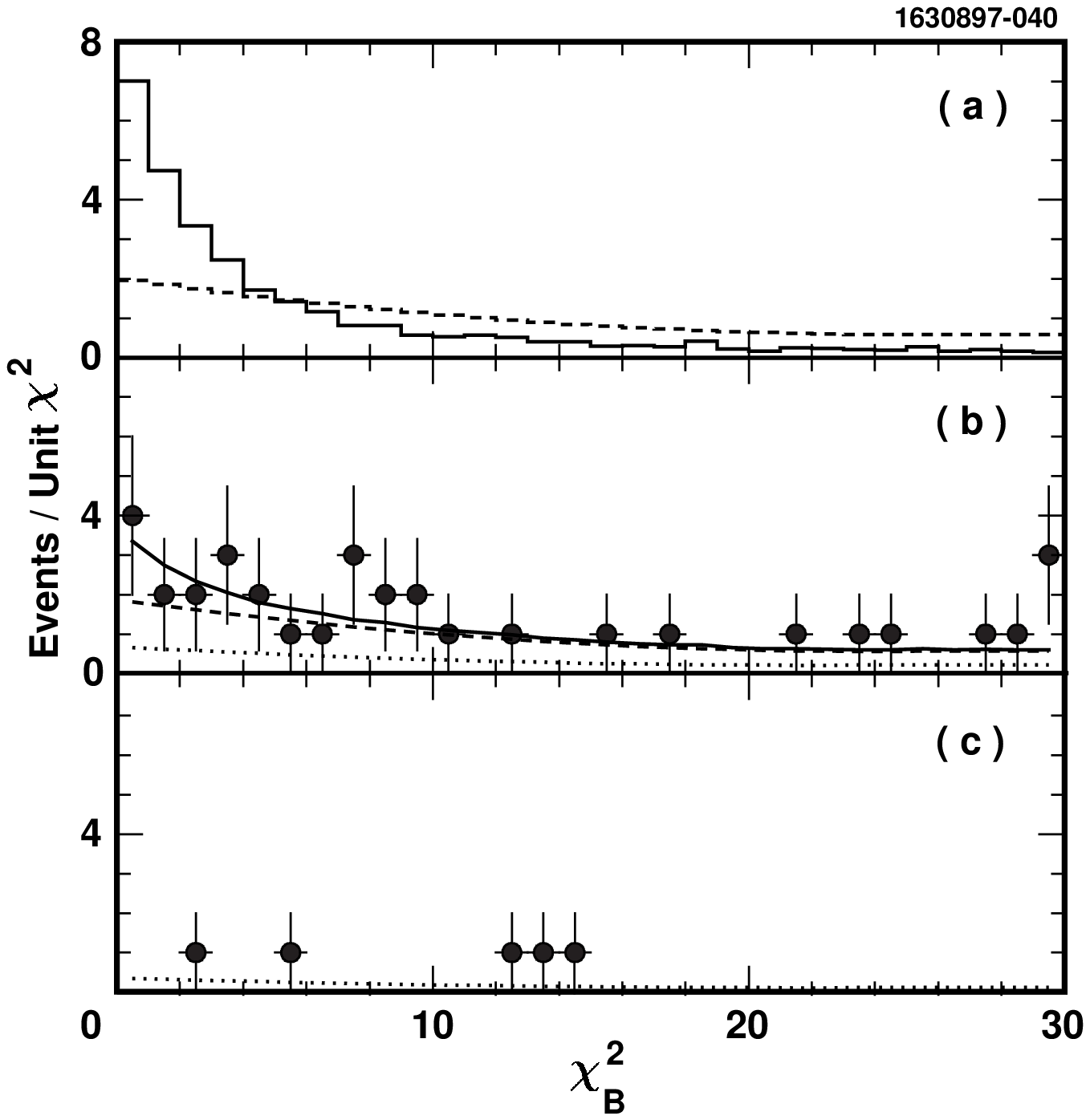,width=\hsize}
 }
 \caption{ 
 $\chi^2_B$ distributions for  $X_s e^+e^-$ data.
 (a) The difference between the expected
 distribution for the signal (solid histogram) and
 $B\bar{B}$ background (dashed histogram).
 Both distributions are normalized to the same area.
 (b) The fit to the on-resonance data (points with error
 bars). The sum of all fitted contributions is indicated by a solid line.
 The fitted 
 background contribution ($B\bar{B}$ plus continuum) is indicated by a dashed
 line. The estimated continuum background, indicated by a dotted  line,
 is simultaneously constrained to the off-resonance
 data (c).
  \label{fig-chib}
 }
\end{figure}

To avoid systematics related to absolute normalization of the $B\bar{B}$
Monte Carlo, instead of counting events after the  $\chi_B^2<6$ cut,
we loosen this cut to $30$ and fit the observed $\chi^2_B$ distributions
in the on- and off-resonance data. We allow for signal contribution,
as well as $B\bar{B}$ and continuum backgrounds. 
The signal is expected to peak sharply at zero, whereas the backgrounds
have flatter distributions (see Fig.~\ref{fig-chib} (a)).
The shapes of all these contributions are fixed from the Monte Carlo
simulation, while the normalizations
are allowed to float. The relative normalization for continuum background
between the on- and off-resonance data is fixed to the known ratio of 
integral luminosities and cross-sections. 
The fitted number of $X_s e^+e^-$,  $X_s\mu^+\mu^-$, 
and $X_s e^{\pm}\mu^{\mp}$ events is
$7\pm7$, $1\pm7$, and $-18\pm10$, respectively. 
As an example, the fit to the $X_s e^+e^-$ data is displayed in 
Fig.~\ref{fig-chib} (b)-(c).

To calculate the signal efficiency and to predict the 
$\chi_B^2$ signal distribution
we generated $b\to s l^+l^-$ Monte Carlo events.
The parton level distributions 
for $b\to s e^+e^-$ and $b\to s \mu^+\mu^-$ 
are predicted from the effective Hamiltonian 
containing Standard Model contributions.
The next-to-leading-order calculations were used \cite{refBM}.
At present, the effect of gluon bremsstrahlung on the outgoing $s-$quark
is only partially included in the theoretical calculations. 
After our $\psi$ and $\psi'$ veto cuts,
the long distance interactions are expected to
constructively interfere with the short distance 
contributions. Estimates of these interference effects
are model dependent. The most recent calculation predicts
modifications of the short distance rate by only about 2\%\cite{Ahmady}\
compared to 20\%\ predicted by some earlier simplified models \cite{otherLD}.
We neglect long distance interactions in our Monte Carlo.
Since no theoretical
calculations for the non-Standard Model decay
$b\rightarrow s e^{\pm} \mu^{\mp}$ exist, we use a phase
space model for these decays.
To account for Fermi motion of the $b$ quark inside the $B$ meson 
we have used the spectator model by Ali {\it et al.} \cite{refAH}. 
The particle content of the $X_s$ system was modeled with the conventional
method of quark hadronization from JETSET \cite{JETSET}.
For better accuracy of the simulations,
when $M(X_s)$ is
in the $K$ or $K^*$ mass region,
the event is regenerated according to the theoretical predictions for
the exclusive $B\to K^{(*)} l^+l^-$ decays by Greub {\it et al.}
\cite{refGreub1}.
The estimated efficiencies are $5.2\%$, $4.5\%$ and $7.3\%$  
for $e^+e^-$, $\mu^+\mu^-$ and $e^{\pm}\mu^{\mp}$ modes respectively.  

To estimate the 
systematic error due to the uncertainty in the $\chi^2_B$ signal and
background shapes, we divide the Monte Carlo sample into low and high
multiplicity channels in the manner 
which produces the largest shape variation.
This shape variation changes the upper limits  by 9\%, 19\%\ and
20\%\ for the $e^+e^-$, $\mu^+\mu^-$ and $e^{\pm}\mu^{\mp}$ 
channels respectively.
Uncertainties in detector simulation are dominated by
charged tracking systematics and are estimated to be 14\%.
Variations of the spectator model parameters \cite{b2sgammaPRL}
result in changes of
the selection efficiency by ($12\pm4$)\%, ($30\pm4$)\%, and ($11\pm4$)\%, 
respectively.
The larger uncertainty in the $\mu^+\mu^-$ channel is the result of 
the lack of muon identification for $P_{\mu}<1$ GeV/c.
Uncertainty in the modeling of the hadronization of the $X_s$ system gives a 
contribution of 9\%.
Adding these sources of systematic errors in quadrature, we estimate
the total systematic errors to be 22\%, 39\% and 28\%,
respectively. 

Using a Gaussian likelihood integrated over positive signal values,
we find upper limits using statistical errors only.
We then loosen these limits by one unit of systematic uncertainty. 

The final results are  
${\cal B} (b \to s e^+ e^-) < 5.7\times 10^{-5}$,
${\cal B} (b \to s \mu^+ \mu^-) < 5.8\times 10^{-5}$,
and
${\cal B} (b \to s e^{\pm}\mu^{\mp} ) < 2.2\times 10^{-5}$.
The results are consistent with the 
Standard Model predictions \cite{refAH},
$(0.8\pm0.2)\times 10^{-5}$, 
$(0.6\pm0.1)\times 10^{-5}$, 
and 0, respectively.
Combining the $e^+e^-$ and $\mu^+ \mu^-$ results,
we also set a limit on the rate averaged over lepton flavors,
${\cal B} (b \to s l^+ l^-) \equiv 
({\cal B} (b \to s e^+ e^-) + {\cal B} (b \to s \mu^+ \mu^-))/2
< 4.2\times 10^{-5}$  (90 \%\ C.L.).

The limit on ${\cal B} (b \to s e^+ e^-)$ 
is more than an order 
of magnitude more restrictive than the previous
limits.
The limit on ${\cal B} (b \to s \mu^+ \mu^-)$ is also
significantly tighter than the
UA1 limit after correcting for the efficiency problem (see discussion 
above).
Furthermore, in contrast with the UA1 measurement, the present
analysis is sensitive to a much wider range of $M(l^+l^-)$.
Therefore, 
our extrapolation to the full phase space is more reliable and
we are sensitive to a broader range of processes beyond
the Standard Model.

We gratefully acknowledge the effort of the CESR staff in providing us with
excellent luminosity and running conditions.
This work was supported by 
the National Science Foundation,
the U.S. Department of Energy,
the Heisenberg Foundation,  
the Alexander von Humboldt Stiftung,
the Natural Sciences and Engineering Research Council of Canada,
and the A.P. Sloan Foundation.


\begin{thebibliography}{99}
\bibitem{UA1sim}
To simulate the 
UA1 acceptance, we use the PYTHIA program \cite{JETSET}
to generate 
$p\bar{p}\to b\bar{b} X$ at 630 GeV in the center of mass energy.
The simulation of $b\to s \mu^+\mu^-$ decays is described later
in the text. 
The acceptance due to the $P_t$ and $M(\mu^+\mu^-)$ cuts 
is already a factor of three lower than the overall
efficiency quoted in the UA1 publication \cite{UA1}.
Further losses are expected
due to inefficiencies in 
the trigger and reconstruction.
\end{thebibliography}
\end{document}